\documentclass[preprint]{IEEE/vgtc}               

\ifpdf
  \pdfoutput=1\relax                   
  \usepackage{graphicx}                
  \DeclareGraphicsExtensions{.pdf,.png,.jpg,.jpeg} 
\else
  \usepackage{graphicx}                
  \DeclareGraphicsExtensions{.eps}     
\fi%

\graphicspath{{figures/}{pictures/}{images/}{./}} 

\usepackage{microtype}                 
\PassOptionsToPackage{warn}{textcomp}  
\usepackage{textcomp}                  
\usepackage{mathptmx}                  
\usepackage{times}                     
\usepackage{cite}                      
\usepackage{tabu}                      
\usepackage{booktabs}                  

\usepackage{xcolor}
\usepackage{marvosym}
\usepackage{fontawesome}
\usepackage{soul}
\usepackage{xurl}
\usepackage{amssymb}
\usepackage{underscore}
\usepackage{enumitem}
\setlist[enumerate]{noitemsep, topsep=1pt}

\newenvironment{packedenum}{
\begin{enumerate}
  \setlength{\itemsep}{1pt}
  \setlength{\parskip}{0pt}
  \setlength{\parsep}{0pt}
  \setlength{\topsep}{0pt}
}{\end{enumerate}}

\onlineid{1133}

\vgtccategory{Research}

\vgtcinsertpkg

\title{Reimagining \textit{TaxiVis} through an Immersive Space-Time Cube\\metaphor and reflecting on potential benefits of Immersive\\Analytics for urban data exploration}

\author{Jorge Wagner\thanks{e-mail: jorge.wagner@ufrgs.br}\\ %
        \parbox{1.3in}{\scriptsize \centering Federal University of\\Rio Grande do Sul}%
\and Claudio T. Silva\thanks{e-mail: csilva@nyu.edu}\\ %
    \parbox{1.3in}{\scriptsize \centering New York University}%
     \and Wolfgang Stuerzlinger\thanks{e-mail: w.s@sfu.ca}\\ %
     \parbox{1.3in}{\scriptsize \centering Simon Fraser University}%
\and Luciana Nedel\thanks{e-mail: nedel@inf.ufrgs.br}\\ %
        \parbox{1.3in}{\scriptsize \centering Federal University of\\Rio Grande do Sul}%
     }

\teaser{
  \centering
	\includegraphics[width=\linewidth]{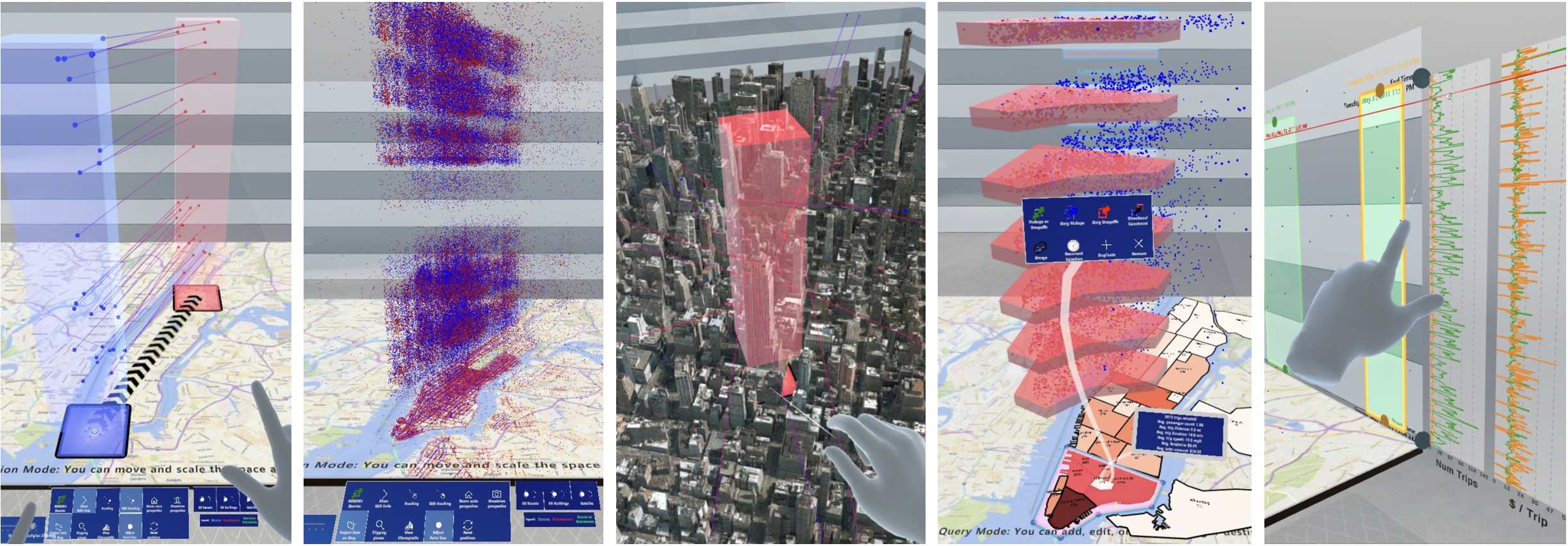}
    \caption{What would \textit{TaxiVis}, a landmark Urban Visualization system from ten years ago, look like if it were to be created within the context of an Immersive Analytics interface? In our reimagination, taxi trips are visualized over space and time (center left), and the visual query language extends to time through 3D prisms (center right). Bi-manual interactions allow brushing and pairwise comparisons (left, center), while embedded time series views bring associated data into the main visualization (right). \href{https://www.youtube.com/watch?v=5nQFVHqUBaU}{\faPlayCircle}}
	\label{fig:teaser}
}

\abstract{
Current visualization research has identified the potential of more immersive settings for data exploration, leveraging VR and AR technologies. To explore how a traditional visualization system could be adapted into an immersive framework, and how it could benefit from this, we decided to revisit a landmark paper presented ten years ago at IEEE VIS. \textit{TaxiVis}, by Ferreira et al., enabled interactive spatio-temporal querying of a large dataset of taxi trips in New York City. Here, we reimagine how \textit{TaxiVis}’ functionalities could be implemented and extended in a 3D immersive environment. Among the unique features we identify as being enabled by the \textit{Immersive TaxiVis} prototype are alternative uses of the additional visual dimension, a fully visual 3D spatio-temporal query framework, and the opportunity to explore the data at different scales and frames of reference. By revisiting the case studies from the original paper, we demonstrate workflows that can benefit from this immersive perspective. Through reporting on our experience, and on the vision and reasoning behind our design decisions, we hope to contribute to the debate on how conventional and immersive visualization paradigms can complement each other and on how the exploration of urban datasets can be facilitated in the coming years.
} 

\CCScatlist{
  \CCScatTwelve{Human-centered computing}{Visu\-al\-iza\-tion}{Visu\-al\-iza\-tion systems and tools}{};
  \CCScatTwelve{Human-centered computing}{Hu\-man computer interaction (HCI)}{Interaction paradigms}{Virtual reality}.
}

\begin{document}


\firstsection{Introduction}

\maketitle

When Ferreira et al. introduced \textit{TaxiVis} in 2013 \cite{ferreira2013visual}, they amazed data experts with the potential provided by a clever visualization tool combining efficient data storage and intuitive real-time visual querying \cite{vo2014pydata}. Over the following decade, their work became highly influential, 
receiving the 10-year VAST \textit{Test of Time} Award at IEEE VIS 2023 \cite{testoftime@VIS2023}. Since then, several research and commercial visualization systems for the exploration of big spatio-temporal origin-destination (OD) urban datasets have been presented, sometimes adopting similar visual querying metaphors \cite{uber2018kepler}. 

In this work, our goal is to reimagine a tool like \textit{TaxiVis} through the lens of \textit{Immersive Analytics} (IA), a novel visualization paradigm that has similarly gained relevance in the past decade \cite{immersiveanalyticsbook, kraus2022survey, fonnet2019survey}. Our idea is to reflect, simultaneously, on (1) how a traditional visualization tool could be implemented in an immersive 3D virtual environment without losing its core functionality and (2) how this scenario could potentially bring in new functionality (what we refer to as \textit{immersive extensions}). We also want to shed light onto potential benefits of IA for the exploration of urban datasets, a relatively under-explored topic despite the 3D nature of cities. 

The background to our work is the growing research on IA \cite{ens2021grandchallenges}, which has shown that the application of immersive AR/VR displays and interactions can improve data exploration activities, for example by offering a new visual dimension \cite{WagnerFilho2018VR}, by supporting different interaction metaphors \cite{cordeil2017imaxes, batch2019spoon} and perspectives \cite{wagner2022frames}, or by facilitating collaboration \cite{ens2021uplift, lee2021shared}. Increasing interest within the VIS, HCI, and AR/VR communities \cite{fonnet2019survey, kraus2022survey} suggest a growing acceptability of such approaches, and a potential partial shift into this direction within visualization research and industry \cite{wiggers2023virtualitics}. The idea is most frequently to complement established techniques, not to replace them \cite{saffo2023shoes, tong2023asymmetric}. While most work at this point is exploratory and not intended for immediate adoption, it is conceivable that such approaches will become more popular over the next years, either due to the release of new immersive devices, due to the discovery of new objective benefits, or simply due to growing consumer and corporate interest. 

In this context, and given the complexity and multidimensional nature of urban data together with the strong requirement for collaboration between people from different backgrounds \cite{ens2021uplift}, we believe that urban data will be a natural application area for IA. Large datasets obtained from cities capture their behavior \cite{doraiswamy2018spatio} and integrate both spatial and abstract information. Alternative interface paradigms such as IA may help analysts, researchers, and citizens leverage such data to better understand human mobility and other urban phenomena in the future. 
Despite the known drawbacks of desktop-based 3D visualizations, the spatial nature of urban spaces has already led to the development of multiple such tools \cite{ferreira2015urbane}, though their adaptation to immersive spaces has seen limited study \cite{chen2017immersiveurban, zhang2021urbanvr}. While some work has targeted immersive movement visualizations \cite{hurter2018fiberclay,yang2018origin}, these have mostly not focused on urban contexts. This might be partially related to the difficulty of the necessary integration of 2D and 3D views and corresponding filters in this domain, one of the issues we seek to address. 

Our vision is that, in the medium-term future, immersive tools will support urban experts in the execution of specific tasks and also enhance remote collaboration. In the long term, immersive AR/VR tools will seamlessly complement existing tools in the workplace, including support for the real-time analysis of data (for example, in a municipal control room during a large urban emergency). 

In this work, we focus specifically on the immersive visualization of origin-destination data, prevalent in the urban context. Besides taxis, multiple other urban systems can provide this type of data, such as ride-sharing apps \cite{uber2018kepler}, 
bike-sharing stations \cite{oliveira2016visual}, transportation cards \cite{zeng2014visualizing}, or mobile phone antennas \cite{reades2007cellular}. We illustrate our ideas through a series of examples employing the taxi trips dataset from New York City (NYC) \cite{donovan2014new}, 
as originally visualized by \textit{TaxiVis}.

To reflect on the potential benefits of immersive visualizations for the exploration of this dataset, we iteratively developed a single-user proof-of-concept prototype. This process helped us envision a series of \textit{immersive extensions} enabled by IA (\autoref{sec:taxivis}), some of which we implemented and others we discuss as future work. These immersive extensions include extending the \textit{TaxiVis}' visual query model to encompass the temporal dimension through a Space-Time Cube (STC) metaphor. The STC view, aided by on-demand 2D projections and embedded time series views, makes temporal patterns in the data more evident and enables the use of visual spatio-temporal filters and brushes through 3D prisms. Other extensions envisioned and already implemented in our prototype include the use of complementary ego- and exocentric immersive perspectives and uses of bi-manual interactions for brushing and comparisons. We also reflect on how to adapt key capabilities of the classic \textit{TaxiVis} system into IA, for example, through the use of embedded views and controls (\autoref{sec:taxivis:adapting}).

In summary, we present as contributions:

\begin{packedenum}
    \item An extensive discussion of the potential benefits and extensions enabled by Immersive Analytics for the urban domain, as well as how an existing visualization tool could potentially be migrated to an immersive paradigm. 
    \item A single-user prototypical system demonstrating most of the ideas we discuss and how they can be implemented, openly available to inform the development of similar applications.\footnote{\texttt{\url{https://github.com/jorge-wagner/ImmersiveTaxiVis}}}
    \item A series of examples that revisit case studies from the original paper and demonstrate the use of our prototype system for the exploration of large real-world urban datasets with OD data. 
    \item From an engineering point of view, a demonstration of how the core ideas behind the \textit{Immersive Analytics Toolkit (IATK)} \cite{cordeil2019iatk} can be adapted to construct new custom applications scalable to reasonably large datasets. 
\end{packedenum}

Further, we also make an effort to guide interested readers to additional relevant work within the IA literature.

\section{Background and Related work} 
\label{sec:related}

Our research is at the intersection of Urban Data Visualization \cite{deng2023survey} and Immersive Analytics \cite{immersiveanalyticsbook}, looking at the benefits stemming from merging these two areas. Within these two areas, we look at relevant work associated with the visualization of urban data, especially origin-destination data, which inspired and informed our design rationale for our reimagined \textit{TaxiVis}.

\subsection{\textit{TaxiVis} and subsequent systems} 
\label{sec:related:urban}

Urban Visualization applications offer analysts visual tools to explore datasets generated by urban spaces, and have gained increased relevance with the progressive release of large open urban datasets. Being able to explore such datasets is becoming highly relevant to better inform expert decisions \cite{doraiswamy2018spatio}.  

\begin{figure}[t]
    \centering
    \includegraphics[height=8cm]{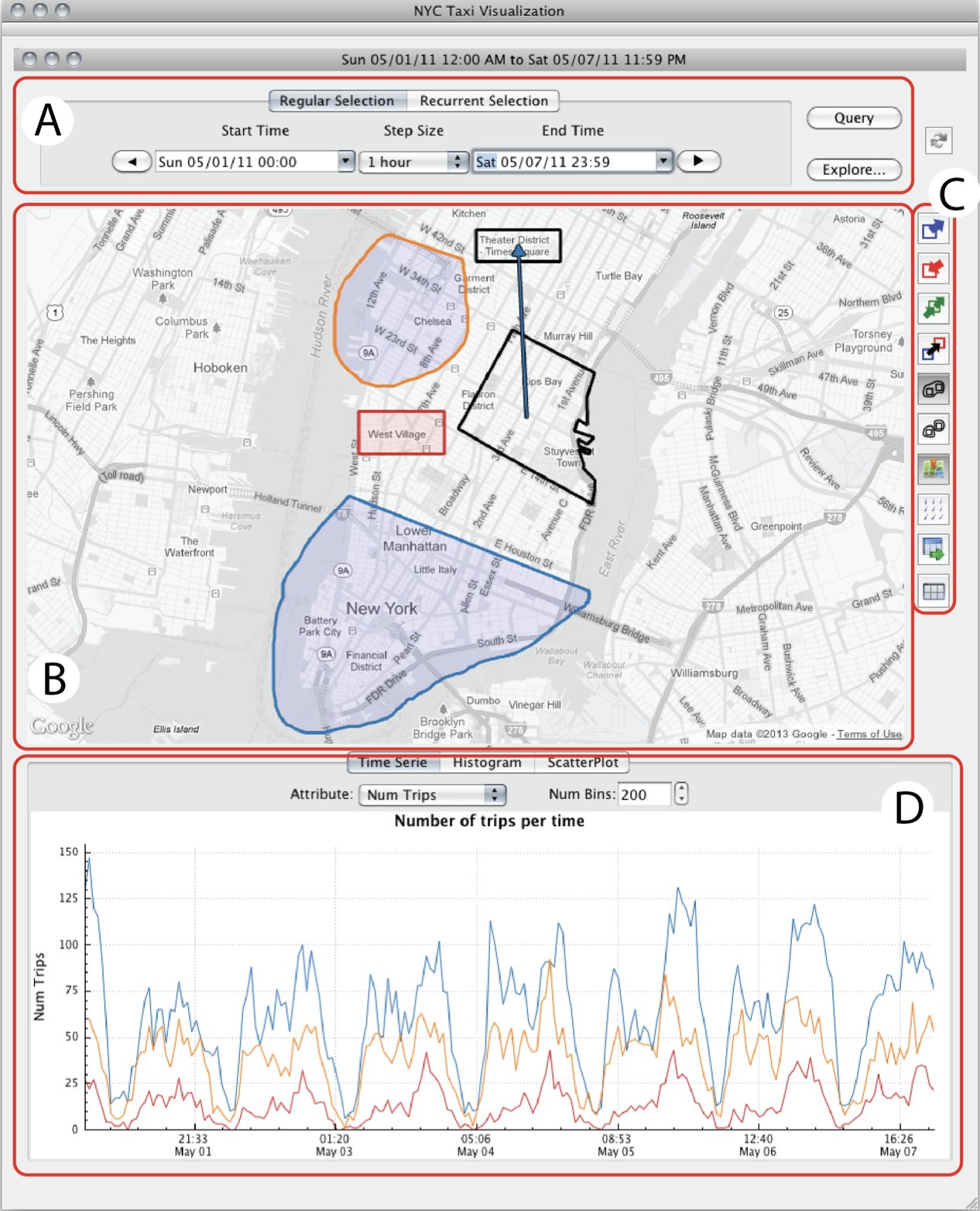}
    \caption{Screenshot of the original \textit{TaxiVis} system illustrating the application of temporal constraints (A) and of visually-defined spatial queries (B). Coordinated views depict associated trip data for each query over time (D). \textit{Source: Ferreira et al. \cite{ferreira2013visual} (\textcopyright2013 IEEE).}}
    \label{fig:rel:taxivis}
\end{figure}

\textit{TaxiVis} \cite{ferreira2013visual} sought to address some of the main challenges in this area at its time of publication, in particular, the difficulty in interactively querying the spatio-temporal urban data. To this end, \textit{TaxiVis} proposed an optimized storage solution and a visual query language to intuitively specify origin-destination queries by drawing or selecting regions on a map. The system also allowed the construction of more complex queries by combining a series of atomic queries and offered a series of coordinated visualizations (see \autoref{fig:rel:taxivis}). The core ideas of \textit{TaxiVis} became highly influential in the field, resulting in a 10-year \textit{Test of Time} recognition awarded at IEEE VIS 2023 \cite{testoftime@VIS2023}.

Subsequent systems have targeted the efficient exploration and uncovering of spatio-temporal patterns in other related large origin-destination datasets, such as bike-sharing data from New York City \cite{oliveira2016visual, shi2019exploring} and subway trips data from Singapore \cite{zeng2016visualizing, yu2015iviztrans}.

\textit{Kepler.gl}, a web-based open-source geospatial visualization system introduced by Uber in 2018 \cite{uber2018kepler} is currently a state-of-the-art tool for the visualization of large origin-destination datasets. It allows the visualization of either raw point or aggregate data through several possible 2D and 3D encodings. Similarly to \textit{TaxiVis}, \textit{Kepler.gl} supports polygon-based visual spatial querying of the data, and also allows interactive origin-destination brushing.

In this work, we reimagine how many of the features introduced by these systems could be extended to an immersive paradigm.

\subsection{Uses of 3D in conventional urban visualization}

Given the inherently 3D nature of urban spaces, the use of 3D visualization approaches in this domain precedes the advent of immersive visualizations. In \textit{Urbane}, for example, Ferreira et al. employed 3D city representations in a desktop-based system to visually depict the effect of planned construction on landmark visibility and sunlight exposure \cite{ferreira2015urbane}. Mota et al. \cite{mota2023urban} compared different approaches to visualize time-varying data embedded in or alongside 3D city models, identifying that plot-based visualizations were more accurate than color-based ones. Moreira et al. \cite{moreira2023urban} introduced \textit{UrbanTK}, a grammar-based toolkit that allows integrating multiple urban datasets into a 3D city model. In our work, we identify better understanding and interacting with the 3D topography of cities as one of the extensions enabled by IA for such applications.

When visualizing spatio-temporal urban datasets, however, another key use of the third dimension is not linked to the 3D representation of cities. As originally proposed by Hägerstraand \cite{hagerstraand1970people}, and revisited by Kraak \cite{kraak2003space}, the third axis can be used to depict the progression of time, resulting in a \textit{Space-Time Cube} (STC). This representation emphasizes the perception of patterns over space and time simultaneously, and has been widely applied in different areas.  

Lock et al. \cite{lock2021sydney}, for example, constructed a 3D STC of one year of bus trips in the city of Sydney, making the occurrence of major delays very evident by unusual point densities. They called this representation a \textit{Performance Point Cloud}. In \textit{iVizTRANS}, Yu et al. \cite{yu2015iviztrans} similarly used an STC view to represent \textit{tap-in tap-out} smart card data from the transport system in Singapore. They rendered each trip between two selected regions as a line connecting origin and destination, highlighting temporal commuting patterns. Cheng et al. \cite{cheng2013exploratory} employed 3D isosurface volumes in STC views to represent the incidence of traffic congestion over time.

Gonçalves et al. \cite{goncalves2015stctaxi}, Ding et al. \cite{ding2016visual}, and Yang et al. \cite{yang2017scalable} all used STC views to visualize, over time, taxi trip \textit{pick-up} and \textit{drop-off} data, for New York, Shanghai, and Shenzhen, respectively. In our work, we integrate similar STC point cloud views of taxi data in the immersive paradigm, where, according to previous findings \cite{wagnerfilho2019stc}, STC exploration leads to lower user mental workload. We also extend the STC metaphor by introducing embedded 2D plots and a 3D spatio-temporal visual query language adapted from \textit{TaxiVis}.

\subsection{Immersive Analytics of movement data and initial attempts at immersive urban analytics}
\label{sec:related:ia}

Started around 2015 in its current form \cite{chandler2015immersive}, Immersive Analytics (IA) is a relatively novel research area at the intersection of Visualization, AR/VR, and HCI. It focuses on facilitating data analysis and understanding through the application of novel visualization and interaction paradigms. The area has been gaining growing prominence in recent years \cite{fonnet2019survey,kraus2022survey}, and current research challenges \cite{ens2021grandchallenges} include supporting the visualization of large datasets \cite{hurter2018fiberclay, cordeil2019iatk}, conducting evaluations with experts and in-the-wild \cite{batch2019spoon, ens2021uplift}, and supporting collaboration \cite{lee2021shared, saffo2023shoes}.

Of particular interest to our current project are the applications of IA to movement data and urban data. 
Notable works in this area include Yang et al's evaluation of immersive origin-destination flow maps at a global level \cite{yang2018origin}. They found that a 3D globe with raised flows was faster and more accurate than other (flat) alternatives. Wagner et al. adapted the previously-mentioned Space-Time Cube metaphor to immersive environments, and, in a comparative study against a desktop-based baseline, found that immersion decreases user workload \cite{wagnerfilho2019stc}. They also identified complementary benefits for egocentric and exocentric immersive perspectives of the cube \cite{wagner2022frames}. In \textit{GeoGate}, Ssin et al. combined an AR STC view with a tabletop display, using a ring-shaped tangible controller to select focal regions on the map \cite{ssin2019geogate}. With \textit{FiberClay}, Hurter et al. extended IA to much larger datasets, combining a custom rendering engine and a series of interaction metaphors to allow the interactive querying of millions of aircraft trajectories. In particular, their system allowed ``sculpting'' queries by manually adding or subtracting selections, and bi-manually brushing combinations of origin and destination areas. All these efforts influenced the design of our \textit{Immersive TaxiVis}.

Other work has also presented initial attempts at applying IA to urban data visualization. Chen at al. \cite{chen2017immersiveurban} presented an early exploration of the design space of immersive urban analytics, proposing, in particular, a typology of visual integration of 3D physical and 2D abstract data, distinguishing between linked views, embedded views, and mixed views. They also proposed an exploded views method to overcome occlusion in urban visualizations \cite{chen2017immersive}.

In recent years, a series of prototypes leveraging different technologies have been designed to integrate data into urban representations. 
In the \textit{CityScope} interactive simulation tool for urban design, Alonso et al. explored a combination of a tangible \textit{LEGO}-based city model overlaid with projected data and a conventional display for 2D coordinated views \cite{alonso2018cityscope}. In \textit{HoloCity}, Lock et al. presented an AR prototype visualizing real-time and aggregate city data, including transportation and social media data, overlaid onto a city model \cite{lock2019holocity}. 

Several works adopted VR approaches. 
In \textit{UrbanVR}, Zhang et al. implemented an immersive system to support the assessment of the impact of new planned buildings \cite{zhang2021urbanvr}, similarly to the aforementioned \textit{Urbane}. 
In \textit{ViBe} (Virtual Berlin), Al Bondakji et al. presented a prototypical 3D environment for sustainable city planning through the visualization of open source datasets of demographic, environmental, and land use data \cite{werner2019vibe}. Spur et al. demonstrated two different metaphors to visualize layers of urban data in VR: stacking several layers at different planes \cite{spur2020exploring} or distorting the data layers into a city-scale \textit{Urban Datasphere} \cite{spur2018datasphere}. Chen et al. proposed a related distortion-based \textit{focus+context} approach, by folding the more distant parts of the map upwards with \textit{UrbanRama} \cite{chen2021urbanrama}.

Further, Cunningham et al. extended the \textit{ImAxes} system \cite{cordeil2017imaxes} to enable embodied interaction with geospatial data, using energy consumption data as an example \cite{cunningham2021geospatial} -- similarly to \textit{ImAxes}, different ways of attaching different data axes then results in different 3D spatial visualizations of the data. Elvezio et al. demonstrated a collaborative prototype for the situated visualization of live urban data such as service requests obtained from public city APIs \cite{elvezio2018urban}, overlaid onto a city model to indicate their origins.

In this work, our primary focus is not on conceiving new applications for immersive urban analytics, but on imagining how a conventional tool could be adapted and extended into the IA paradigm. To this end, all examples above informed our proposed designs.

\section{Envisioned Immersive Extensions}
\label{sec:taxivis}

In this section, we detail how we envision that IA could potentially extend the functionality of \textit{TaxiVis}, introducing a series of \ul{\textit{immersive extensions}} enabled by IA (underlined throughout this section). In the next section, we will focus on how it could also adapt features originally designed for a desktop interaction paradigm. 

We start with an expansion to the \textit{TaxiVis} visual query model by integrating the temporal component into the visual representation, following the Space-Time Cube (STC) metaphor \cite{bach2014review, wagnerfilho2019stc}. Further extensions correspond to alternative uses of the third dimension for: immersive interaction, immersive perspectives, immersive collaboration, and immersive engagement. This conceptual vision was iteratively informed by the development of our \textit{Immersive TaxiVis} proof-of-concept prototype, which we start introducing in this section to illustrate the discussed concepts. Extensions already implemented into this prototype are marked with check marks (\checkmark).

We implemented this prototype using the Unity game engine, the \textit{Immersive Analytics Toolkit (IATK)} \cite{cordeil2019iatk} for data rendering, the \textit{Mixed Reality Toolkit (MRTK)} for interaction support, and the \textit{Bing Maps SDK} for mapping resources. Please refer to the Supplementary Materials (\autoref{sec:taxivis:implementation}) for more information on how we adapted the core ideas of \textit{IATK} to implement spatio-temporal queries.

\subsection{Extended visual query model (or: \textit{TaxiVis} meets the Space-Time Cube)}
\label{sec:taxivis:stc}

The additional visual dimension enabled by an immersive visualization approach, i.e., the third dimension, can be utilized in many different ways when working with urban data, as we will discuss later (\autoref{sec:ext:3d}). Our main idea to leverage this additional dimension for the exploration of urban origin-destination data is to \ul{extend the visualization to encompass the temporal component \checkmark}, through a Space-Time Cube (STC) metaphor. In an STC, the vertical axis corresponds to time (which advances upwards in all examples in this paper, although the opposite is also possible \cite{wagnerfilho2019stc}), enabling a clear visualization of patterns over space and time. While the original \textit{TaxiVis} system offered the possibility of visualizing several time steps side-by-side, an STC could be seen as a potentially infinite number of time steps in the same view. Our examples in \autoref{sec:demo} illustrate how this can be advantageous, for example, to make temporal patterns more evident. As discussed earlier, this approach has already been explored non-immersively in recent work \cite{lock2021sydney, ding2016visual, yang2017scalable, goncalves2015stctaxi}. Prior work also demonstrated that an immersive STC can minimize the interaction challenges previously associated with this visualization \cite{wagnerfilho2019stc}. 

In an STC for origin-destination data, data positions can be either represented individually, as a point cloud \cite{lock2021sydney}, or connected to each other, forming space-time paths \cite{goncalves2015stctaxi, menin2019estime}. We enable both such visualization forms in our prototype. Following the \textit{TaxiVis} convention, we color \textit{pick-up} points in blue and \textit{drop-off} points in red. Users are able to adjust point sizes through a menu slider. Smaller data points lead to better visualization of patterns in the full point cloud, while bigger data points can help emphasize a selected subset. To avoid clutter, showing connecting lines between pairs is more convenient when associated with specific origin-destination filters, such as in \textit{iVizTRANS} \cite{yu2015iviztrans}. Connected trajectories are also well suited for data with multiple origin-destination pairs for a same individual---in this case, they can also be colored according to thematic attributes, for example, the means of transportation or the trip purpose \cite{menin2019estime}. Showing aggregate views of the STC data is also possible through a density heat map \cite{demvsar2010space}, but currently not implemented. We also support on-demand 2D projections of inspected data points both onto the map and onto the time walls to allow the exploration of each component separately and reduce the effort needed to understand the locations and times of mid-air trajectories.

A key contribution of \textit{TaxiVis} was its \textbf{visual query model}, through which queries could be visually composed by combining spatial, temporal, and attribute constraints, and iteratively refined by direct manipulation of the results. One could manually draw spatial queries on the map view (\autoref{fig:rel:taxivis}--B), while specifying temporal queries in a different module of the interface, the \textit{time selection widget} (\autoref{fig:rel:taxivis}--A). For \textit{Immersive TaxiVis}, we propose \ul{extending the visual query model \checkmark} so that a user can query both spatial and temporal components simultaneously by creating and adjusting visual representations in the STC view. To this end, instead of 2D query polygons, we support query prisms (i.e., extruded polygons) in 3D space, whose height corresponds to their temporal span, as seen in \autoref{fig:teaser}. 

In our prototype, we support all kinds of queries proposed by the original \textit{TaxiVis} model, with the extension of handling the temporal component as inseparable from the spatial one.

\textbf{Atomic queries.} As in \textit{TaxiVis}, atomic queries can be created by either freely drawing a region on the map, a time span on the time walls, or by selecting a specific neighborhood on the choropleth map. A new query specified on the map targets initially the whole time span of the dataset, and vice-versa. Besides being visually represented by the query prisms, queries are also visually projected on the map and on the time walls, to facilitate the user's understanding. Wall projections are also labeled to indicate their lower and upper limits. Users can freely translate and scale the query prisms to modify queries, and they can also directly manipulate the query projections to modify their limits. The time granularity of the time walls is dynamically adjusted according to how much the user zooms in or out on the data, to facilitate selecting periods of interest. As in \textit{TaxiVis}, each atomic query is assigned a new query color, which is used to color the projection contours. This color is used to refer to the query in associated visualizations. 

Atomic queries can indicate origin-only (taxi pick-ups, in the case of taxi data) or destination-only constraints (taxi drop-offs), or their combination (either pick-ups or drop-offs). Following \textit{TaxiVis}'s encoding, these are indicated by blue, red, or green areas or volumes, respectively. To avoid overloading the main menu (see \autoref{sec:taxivis:adapting}), we defined that new queries correspond to both origin and destination (i.e., green volumes) by default. When a new query is created, a 2D panel of buttons (see \autoref{fig:teaser}--\textit{center right}) appears with it, which enables query modifications. 
This panel is dynamically positioned close to the user to facilitate interactions, with a leading line linking it to its query. Through these buttons, the user can modify the kind of constraint it represents. To avoid overloading the interface, these buttons are automatically hidden when the user exits the query editing mode, and the query prism is then also made more translucent (or invisible, depending on user preference). Each query is also connected to a floating 2D panel summarizing statistics about the selected trips.

\textbf{Directional queries.} One of the most semantically useful query types in \textit{TaxiVis} might be the directional one, where the user specifies both origin and destination constraints. Such a query can be used, for example, to select taxis that departed from an airport and went to a given area in Manhattan. In our implementation, the user can select one of the atomic query buttons (\textit{``Directional Constraint''}) to directionally link a query to another existing or new one. Such a link is visually indicated by an arrow line rendered over the map surface, as illustrated in \autoref{fig:teaser}--\textit{left}. A new directional query is immediately assigned a single color and a single set of buttons (where one of the options, naturally, is to revert it to its two atomic queries).

\textbf{Recurrent queries.} In \textit{TaxiVis}, recurrent temporal selections (e.g., every Monday, every day from 10 to 11AM) could be filtered through the time selection widgets. As we extend the visual representation to integrate time, we decided to represent queries associated with recurrent selections as sliced prisms, clearly indicating the currently applied selections, as shown in \autoref{fig:teaser}--\textit{center right}. Such selections can be applied either to one individual query or to all existing queries. The selection widget is designed similarly to the original interface.

\textbf{Merged queries.} Multiple atomic queries can be merged into single \textit{merged query} (or \textit{complex query}) by logical disjunction (\textit{``or''}), so that the results of multiple alternative constraints are aggregated in the associated visualizations. This action can be performed by selecting the \textit{``Merge''} button on a query panel, and then selecting a new query to be merged into it. Merged queries are assigned a single color and a single set of query option buttons.

\subsection{Other potential uses of the third dimension} 
\label{sec:ext:3d}

As discussed above, our main use for the third dimension enabled by IA is to integrate the temporal component into the visualization and thus render all time steps on top of each other through an STC. 

Another obvious use of 3D in the urban context, which we also support (see \autoref{fig:teaser}--\textit{center}), is \ul{correlating the data with the urban topography \checkmark}. Depending on the dataset, features such as flat or hilly areas can be of interest due to their impact on different forms of mobility, as can be the density of tall buildings in a given area. Some desktop-based urban visualization systems, such as \textit{Urbane} \cite{ferreira2015urbane}, already leverage 3D views to visualize building shadows and landmarks visibility. We believe that the combination of the stereoscopic display and more suitable 3D interaction metaphors will increase the utility of the 3D views when compared to conventional systems.

Other possible uses of 3D which are not included in our prototype would include \ul{layering query results} into a limited number of planes or time steps to facilitate comparisons, thus resulting in a \textit{Stratified STC}, which would be comparable to the side-by-side views offered by \textit{TaxiVis}. Depending on the dataset and user needs, we could also have a \textit{Cyclic STC} which resets the data height after a certain amount of time (depending on the desired granularity, such as day or hour) in order to \ul{compare mobility patterns over time}. This latter option would be more convenient for smaller datasets.

Finally, instead of visualizing individual data points, the 3D space can also be used to render \ul{alternative 3D geospatial visualizations}, such as 3D heatmaps, as studied by Kraus et al. \cite{kraus2020heatmaps}, or \textit{Prism Maps}, as studied by Yang et al. \cite{yang2021tilt}.

\subsection{Uses of immersive and bi-manual interactions}

Our system supports interactions with either controllers or hand tracking in the \textit{Oculus Quest 2} headset. 
Both modes support at-a-distance ray-based interactions, and the hand-tracked mode also supports direct touch interactions with menus. This combination \cite{wagner2021combining} enables easy interaction with both 2D and 3D interface components. 

\begin{figure}[h] 
 \includegraphics[width=0.49\linewidth]{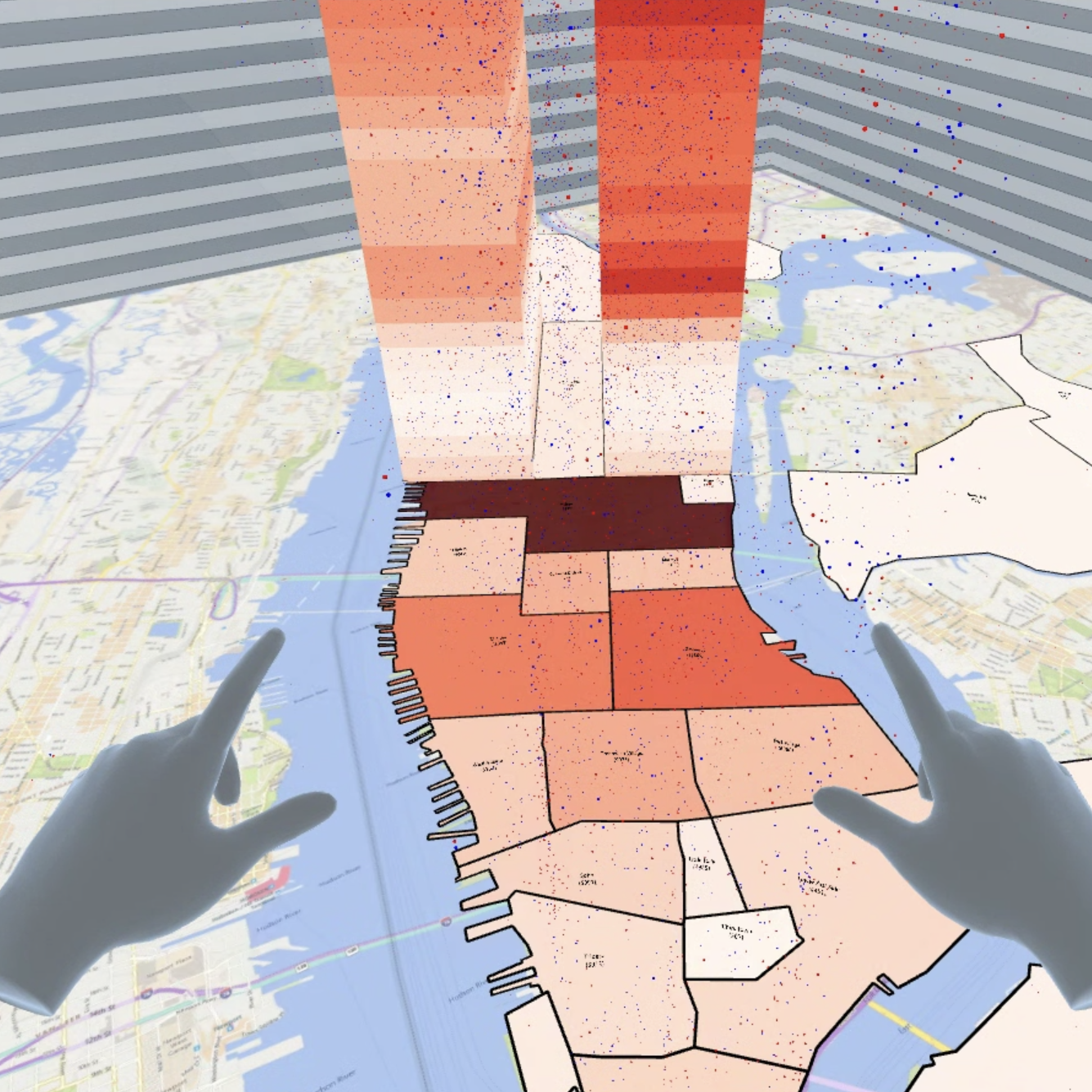}
\hfill	
  \includegraphics[width=0.49\linewidth]{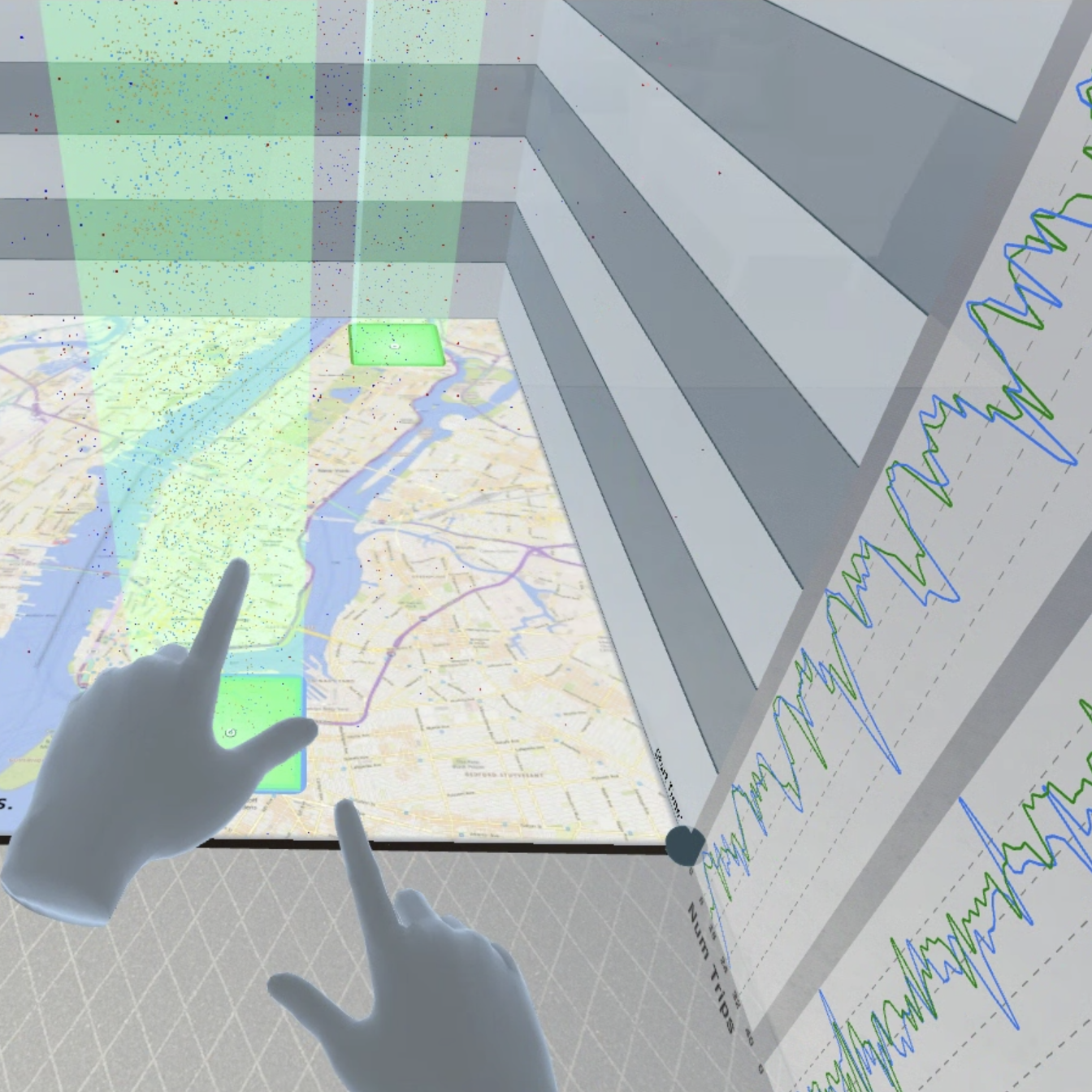}     

    \caption{Extensions enable pairwise comparisons between neighborhoods with bi-manual interaction, either through choropleth prism stacks (left) or two-handed brushes and embedded plots (right).}
	\label{fig:ex:neighborhoods}
\end{figure}

The fact that an immersed user conventionally has access to two interaction devices (controllers or hands) instead of one also enables data exploration by \ul{simultaneous origin-destination brushing \checkmark}, i.e., dynamically specifying both origin and destination constraints by pointing to different locations or time periods with each hand (see \autoref{fig:teaser}--\textit{left}), as proposed originally by \textit{FiberClay} \cite{hurter2018fiberclay}. This could also be seen as an extension of the single desktop-based time lens proposed by \textit{Trajectory Wall} \cite{tominski2012stacking} or the brushing feature available in \textit{Kepler.gl} \cite{uber2018kepler}. 
Two-handed brushing can also be used for \ul{quick pairwise comparisons \checkmark} between two simultaneous selections. We support this through prism brushes (\autoref{fig:ex:neighborhoods}--\textit{right}) and through our \textit{choropleth mode}, where users can quickly compare stacked color-coded prisms for two different neighborhoods, allowing a detailed comparison of traffic patterns over time (\autoref{fig:ex:neighborhoods}--\textit{left}). This latter option is similar to the visualization of energy consumption over time in \textit{Uplift} \cite{ens2021uplift} and \textit{TimeTables} \cite{zhang2022timetables}. We aggregate time dynamically according to the temporal zoom level applied to the STC. Pairwise comparisons are also available through the \textit{cutting planes} (one controlled by each hand) and \textit{finger inspection lenses} (tapping data points to obtain details on demand) features.

\subsection{Complementary uses of immersive perspectives}

\begin{figure}[h] 
 \includegraphics[width=0.49\linewidth]{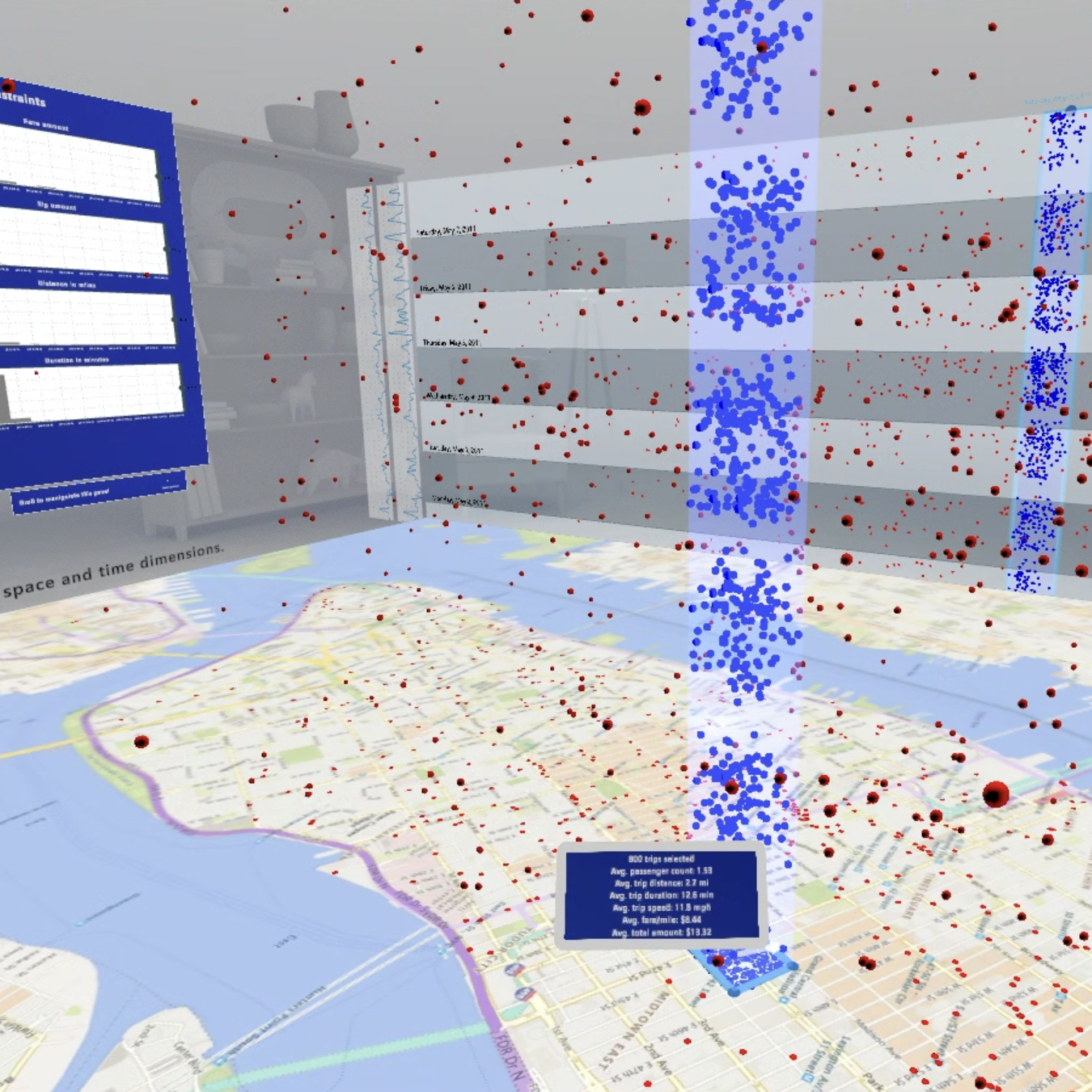}
\hfill	
  \includegraphics[width=0.49\linewidth]{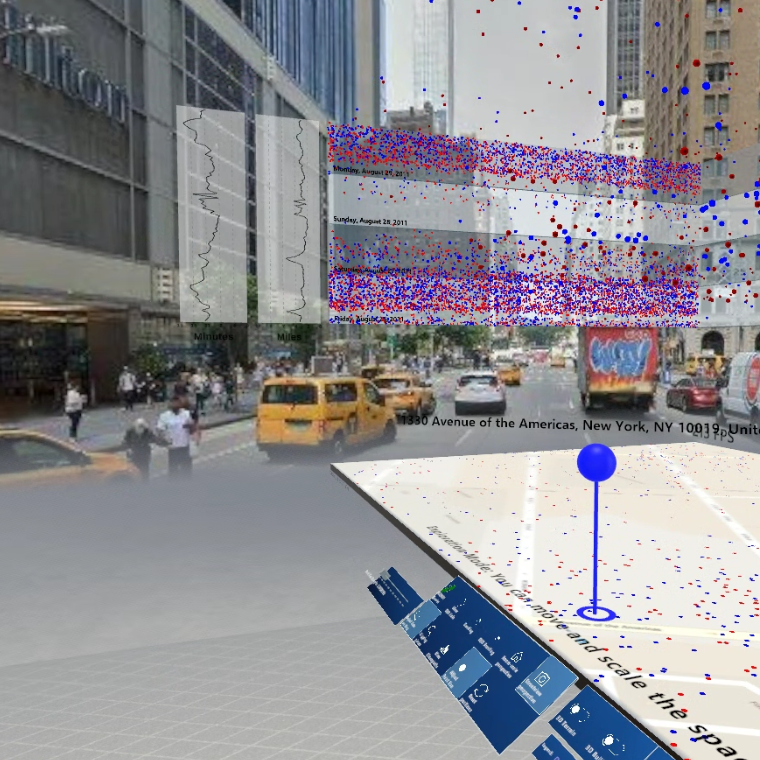}     

    \caption{Immersion allows the integration of complementary frames of reference, such as egocentric room-scale (left) and egocentric view of 360-degree street images (right).}
	\label{fig:egocentric}
\end{figure}

We are convinced that a tabletop metaphor (as seen in \autoref{fig:ex:sandy}) is the ideal standard mode for most urban data exploration scenarios. This metaphor evokes the familiar concept of an architectural scale model and allows easy user operation, replacing most navigation with manipulation actions. Also, it supports a group of collaborating users around the ``table'', such as seen in \textit{Uplift} \cite{ens2021uplift} and \textit{CityScope} \cite{alonso2018cityscope} (where real tabletop screens were used). Depending on user preference, it also supports both seated and standing use. However, as demonstrated by Wagner et al. \cite{wagner2022frames}, both exocentric and egocentric immersive perspectives can be preferable depending on the task and goal. An interesting extension enabled by IA is \ul{the ability to inspect urban data at different perspectives and scales \checkmark}, including egocentrically in room-scale when desired (see \autoref{fig:egocentric}--left).

Another interesting approach enabled by VR in urban spaces is the \ul{inspection of data at real city-scale \checkmark}. The idea is that the user would be able to \textit{dive-in} and \textit{dive-out} of this real-scale egocentric view to inspect the actual location where some phenomena of interest transpired. 
Given the currently limited resolution of photo-realistic 3D tiles such as \textit{Bing}'s at the street level, we opted to implement this functionality by using 360 degrees images from \textit{Google StreetView}, an approach already adopted in some commercial applications such as \textit{Google Earth VR} and \textit{Wooorld}. By positioning a 3D widget, the user can select any position on the map, and the closest possible image is then loaded as a scene \textit{skybox}, preserving the STC view as context (see \autoref{fig:egocentric}--right). In future work, we intend to replace \textit{StreetView} data with a larger dataset of millions of images of NYC over time to allow coverage of the temporal dimension as well \cite{miranda2020urban}.

\subsection{Future uses of immersive collaboration}

Enhanced support for collaboration is a key IA research challenge \cite{ens2021grandchallenges} with multiple potential applications to immersive urban analytics. As demonstrated by \textit{Uplift}, IA may be particularly useful to support knowledge sharing between users with different backgrounds \cite{ens2021uplift}. 
At the present time, supporting such collaboration has not been the focus of our proof-of-concept prototype, but we identify the following possible designs as of particular interest: (1) \ul{remote collaboration} enhanced by avatars and body cues \cite{lee2021shared}, (2) access to both \ul{private and shared views} \cite{reipschlager2020personal}, (3) \ul{multi-scale collaboration} (i.e., between users immersed at different scales to investigate different aspects of the data) \cite{piumsomboon2019shoulder}, and (4) \ul{asymmetric collaboration} (i.e., between immersed and non-immersed users to leverage the strengths of each paradigm) \cite{saffo2023shoes, tong2023asymmetric}.

\subsection{Future uses of immersive engagement}

Besides the potential advantages already discussed, immersive interfaces can be considered engaging and playful to casual users due to their novelty. This can be leveraged to \ul{support and motivate casual data exploration} by general citizens \cite{pousman2007casual}. In this context, recent work has explored immersive urban simulations to support participatory planning \cite{schrom2020planning,keil2023measuring} and urban noise understanding \cite{berger2019sonification}.

One possible future way of integrating a more playful metaphor into \textit{Immersive TaxiVis} could be simulating the movement of taxis in the city at any given time by showing 3D taxi models, which could be implemented either at a city- or desk-scale. This feature would then be comparable to the animations used in \textit{TaxiVis} to illustrate the probable route taken by the visualized trips. Such an interface would also qualify as \textit{data visceralization}, i.e., immersive experiences that facilitate intuitive understanding of quantities \cite{lee2021visceralization}.

\section{Adapting desktop features to IA}
\label{sec:taxivis:adapting}

Beyond extending \textit{TaxiVis}' functionality, as discussed in the previous section, we consider it essential to adapt the original system's functionality effectively to an immersive context. In this section, we discuss the design rationale for our \textit{immersive adaptations} of elements from \textit{TaxiVis}' original WIMP (\textit{Windows, icons, menus, pointer}) interface. To this end, we leveraged, in particular, a combination of embedded views, coordinated views, and contextual menus, as discussed below.

\subsection{Supporting interactions with space and time}

To support the core functionality of \textit{TaxiVis}, our first priority was to allow efficient interactions with space and time. Besides the strong stereopsis depth cues provided by VR headsets, we integrated additional features to facilitate comparisons, particularly in the temporal (vertical) dimension. Inspection lines indicate the date and time currently pointed at by each hand. All points corresponding to the temporal slices in focus are highlighted in a darker color (slices adjust dynamically to the level of temporal zoom applied). Translucent cutting planes are also available on demand, and users can inspect any data point with their fingers through \textit{inspection lenses}. 

To avoid overwhelming the user and leading to unintended interactions, we divided the system into two main interaction modes: \textit{exploration mode}, where the user is free to drag and zoom into space and time, and \textit{query mode}, where the user is free to create new spatio-temporal queries or edit existing ones.

\subsection{Uses of embedded views and controls}

A significant part of \textit{TaxiVis}' original interface is dedicated to temporal constraints (\autoref{fig:rel:taxivis}--A) and visualization of time series data (\autoref{fig:rel:taxivis}--D) for the identification of patterns over time. When distributing the taxi data over space and time simultaneously, our design rationale was to also integrate as much temporal information as possible into this Space-Time Cube view. 

In this context, we implemented the ability to embed all attribute time series views into the STC walls, as shown in \autoref{fig:teaser}--\textit{right}, \autoref{fig:ex:sandy}, and \autoref{fig:ex:airports}. As in \textit{TaxiVis}, time series can display data for the whole dataset or per query, and can be switched between line graphs and histogram views depending on user preference. These views dynamically move and scale with the STC time walls.

Similarly, the controls for \textit{regular selection} of continuous time intervals are also fully embedded into the time walls, as sphere widgets on the right side of the STC (see \autoref{fig:teaser}--\textit{right}). However, we opted to preserve the \textit{recurrent selection} widget as a separate panel, as we considered the user interface would become unnecessarily more complex if integrated into the STC walls. As discussed in the next section, we also preserved a separate 2D panel for time series views in case this view is more familiar and preferable to the user.

Besides embedding the temporal data, we also overlay a choropleth view of the data on the 2D base map of the STC, similarly to \textit{TaxiVis}. As mentioned, this view can be used to create new neighborhood queries and for pairwise comparisons between neighborhoods.

\subsection{Uses of coordinated views}

\begin{figure}[h] 
	\includegraphics[width=0.49\linewidth]{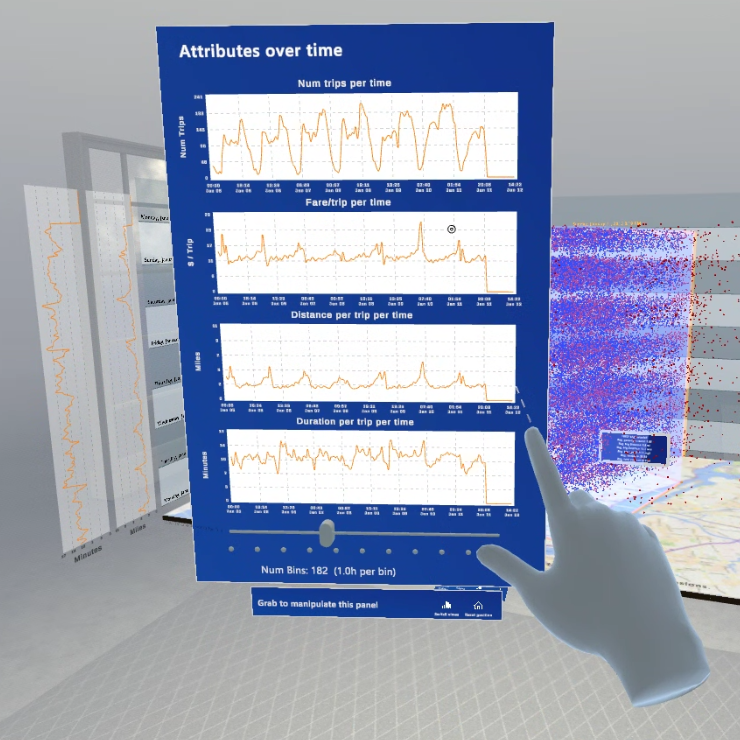}
    \hfill
 	\includegraphics[width=0.49\linewidth]{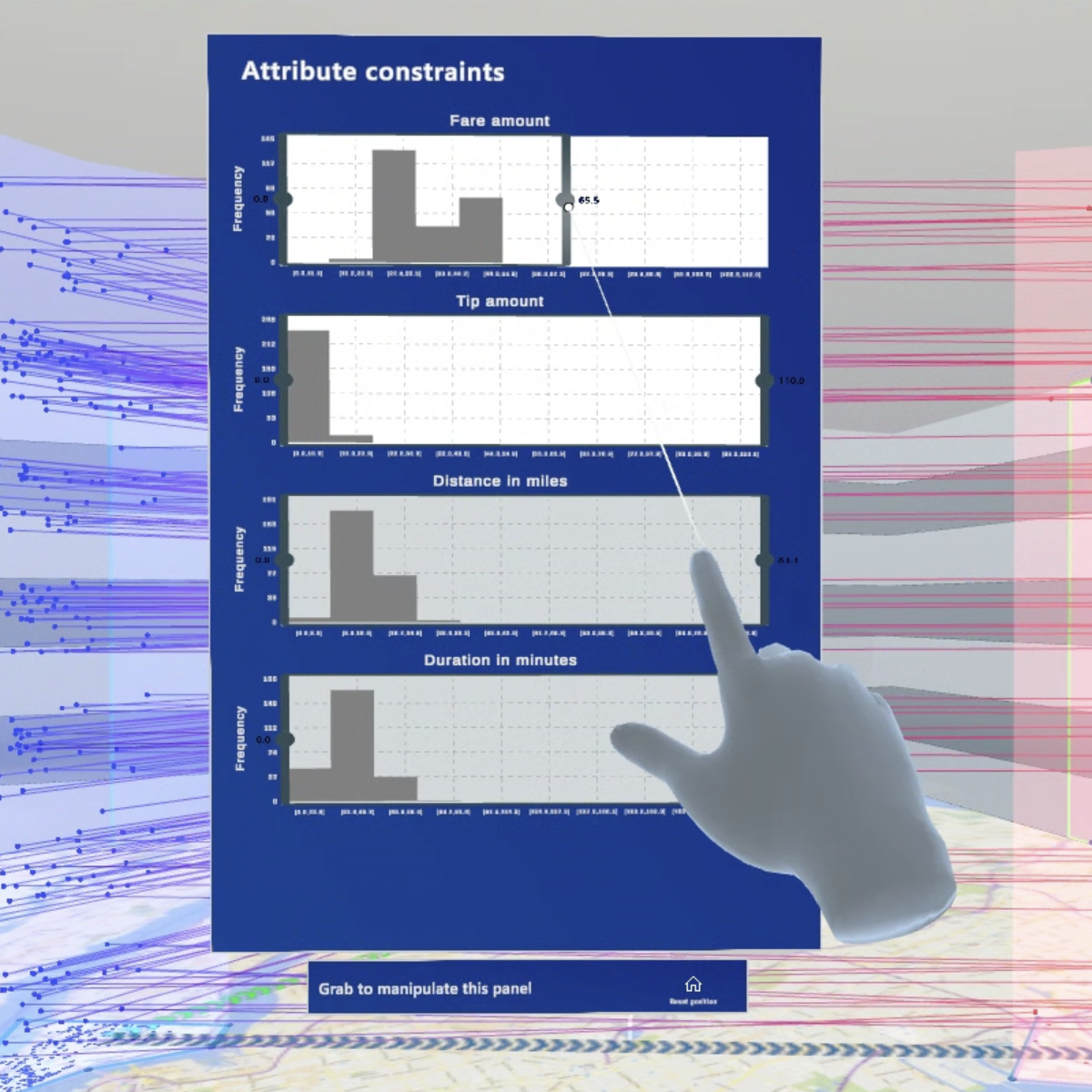}
    \caption{Movable 2D panels complement the interface, offering similar attribute exploration functionality as in \textit{TaxiVis}.}
	\label{fig:coordinated}
\end{figure}

The coordinated views offered in \textit{TaxiVis} through different windows for attribute exploration are also available through separate 2D panels in our prototype (\autoref{fig:coordinated}). These include the \textit{attribute constraints} panel that allows the analysis of value distributions and application of minimum and maximum constraints, and the \textit{attributes over time} panel that displays time series in a more conventional format. 

Leveraging the 3D space of the immersive environment, the coordinated panels sit by default at the sides of the user in the exocentric desk-scale mode, but they are fully movable and can be positioned next to or inside the STC. In room-scale mode, they are originally positioned within an open side of the STC walls, but users can bring them inside the STC if preferred (as in \autoref{fig:coordinated}--\textit{right}).

\subsection{Uses of contextual menus}

\textit{TaxiVis}' toolbar (\autoref{fig:rel:taxivis}--C) provides buttons for the creation and combination of different kinds of queries and for activating different features. Some of these buttons work in combination with conventional mouse-keyboard metaphors, such as holding a key for multiple selections, or with keyboard shortcuts. In our proposed implementation, we also have a main menu of buttons, which can be attached to the user's hands (revealed with a palm-up gesture), to the hand controller, or to the desk edge (see \autoref{fig:teaser}--\textit{left}). Through this menu, users can enter the query editing mode, and activate several features we propose, such as displaying links, displaying cutting planes, displaying point projections, and changing perspectives.

However, taking advantage of the 3D space, we also propose to decompose this menu into contextual menus as much as possible. We believe this avoids overloading the interface and makes relevant buttons easier to find when needed. In this context, map settings (such as activating satellite images or 3D buildings) are attached to the desk, while query modification buttons are attached to each specific query (see \autoref{fig:teaser}--\textit{center right}). To avoid having menus out of user reach, the query menus, which are only visible in the query editing mode, are positioned closer to the user's head.

\begin{figure*}[t]
\centering
    \includegraphics[width=\linewidth]{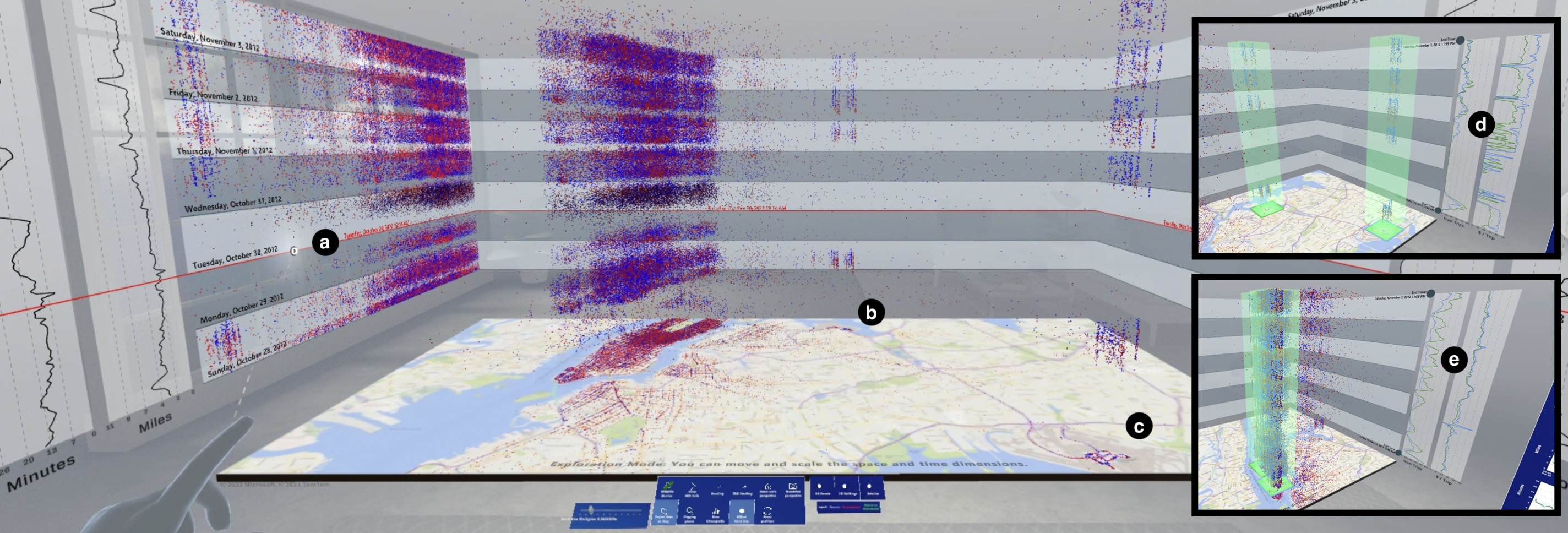}
    \caption{Overview of taxi activity in the week of Hurricane Sandy in October 2012. The STC point cloud and its projections (a) evidence the sudden drop in activity from Monday to Tuesday and the gradual recovery in Lower Manhattan over the next days. The insets highlight the use of interactive brushing and embedded plots to compare activity at different airports (top) and in different parts of Manhattan (bottom).}
	\label{fig:ex:sandy}
\end{figure*}

\section{Concept Demonstration} 
\label{sec:demo}

In this section, we present and discuss a series of examples that showcase how our adapted and extended design works for exploring the NYC taxi dataset.
To reflect on how an immersive reimagination could reproduce and extend \textit{TaxiVis}'s workflows, we revisit case studies from the original paper \cite{ferreira2013visual}. 
As discussed earlier, our idea is not to replace \textit{TaxiVis}, which works remarkably well for what it was designed for, but to reflect on the potential benefits opened up by IA.

We use taxi data from 2010-2015 \cite{donovan2014new} in these examples, as more recent data no longer includes detailed latitude-longitude information due to privacy concerns. The analysis of newer data with approximate locations is still possible but would not be as directly comparable to \textit{TaxiVis}. Given the scalability challenges we discuss in \autoref{sec:disc:impl}, in these examples, we work with randomly sampled subsets of data containing up to a hundred thousand taxi trips each.

\subsection{Using the self-evidence of temporal patterns for an improved analysis of behavior over time}
\label{sec:demo:hurricanes}

The workflow for the identification of unexpected temporal behaviors could start by analyzing a time series distribution of trips over time, using the coordinated 2D panel or the embedded plots (for example, in \autoref{fig:teaser}--\textit{right}, the unusually low activity at the STC's top corresponds to Memorial Day). While this would be typically followed by an inspection of periods of interest through the \textit{time-space exploration mechanism} in \textit{TaxiVis}, i.e., creating a series of small multiple visualizations with a user-defined step size, in our immersive system, the STC point cloud presents a clear picture of the evolution of an event over time, helping users to detect and inspect both gradual or sudden variations of behavior over time. Some features, such as the ability to manually adjust point sizes, allow the user to make patterns more evident. Further, bi-manual cutting planes and brushes also help in identifying periods of interest. On-demand 2D projections on the map and walls (\autoref{fig:ex:sandy}--\textit{a}) also contribute to the analysis when combined with mid-air brushes and queries. 

A good example is the case of major weather events that disrupt life in NYC. Like Ferreira et al., we analyze the cases of Hurricanes Irene and Sandy, which severely impacted the city in August 2011 and October 2012, respectively. \autoref{fig:teaser}--\textit{center left} depicts the evolution of Irene, while \autoref{fig:ex:sandy} depicts Sandy. The STC view allows a clear comparison between the patterns for each storm, evidencing, for example, how Irene led to a more extreme interruption of taxi services. While the same conclusion was reached by Ferreira et al., in our system, it is more straightforward to arrive at this insight.

The STC also helps highlight location-specific temporal patterns. One example is the clear difference between activity at the airports LaGuardia (\autoref{fig:ex:sandy}--\textit{b}) and JFK (\autoref{fig:ex:sandy}--\textit{c}) following Sandy. Clearly, taxis return to JFK on Oct. 31st, while they only return to LaGuardia on Nov. 1st, what is also confirmed by the embedded plots when brushing these areas (\autoref{fig:ex:sandy}--\textit{d}). News articles from the time confirm that LaGuardia flights resumed later due to significant flooding. Another example for the use of comparative brushes or queries is an analysis of how activity was restored in different areas after the storm. \autoref{fig:ex:sandy}--\textit{e} illustrates a comparison of Midtown to Lower Manhattan, with the latter presenting a slower recovery due to a power outage. Moving into a room-scale view of the data and going inside the point cloud is also useful to inspect such patterns. 

Reflecting on future use scenarios, and assuming the use of real-time data, the Space-Time Cube view could also aid an analysis by providing clear contextual information to current activity.

\subsection{Using query prisms and alternative perspectives to support the analysis of transportation hubs}
\label{sec:demo:airports}

Taxi data can also shed light on how people move in and out of the city, by focusing on taxi trips to and from transportation hubs such as airports and train stations. 
In \textit{Immersive TaxiVis}, it is possible to investigate these hubs with the help of pointer brushes (indicating either regions or origin-destination pairs) or query prisms positioned around these locations. The height of the prisms can be adjusted to specify a period of interest, and recurrent selections can focus on specific parts of the day, for example (\autoref{fig:teaser}--\textit{center right}).  

\begin{figure*}[h]
    \centering
    \includegraphics[width=\linewidth]{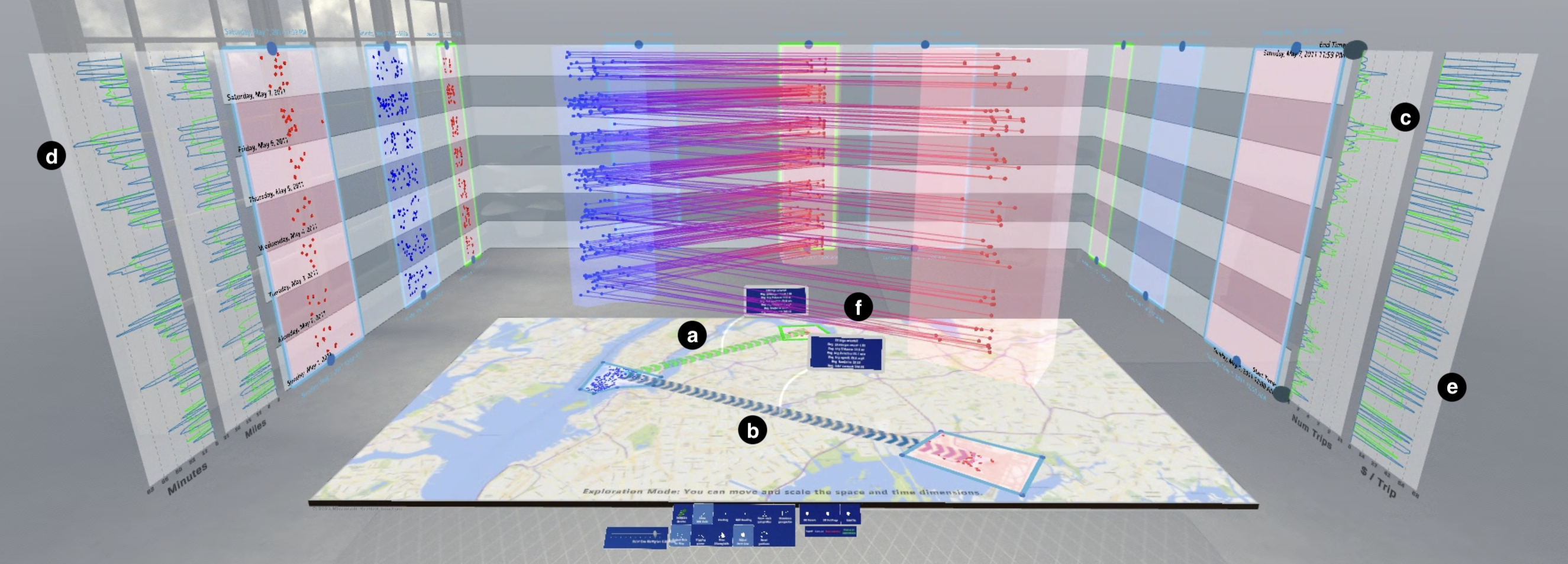}
    \caption{Directional queries selecting trips departing from Lower Manhattan to LaGuardia (a) and JFK (b) airports. The embedded plots indicate how trip numbers (c) and duration (d) vary during one week. On-demand origin-destination connecting lines (f) help emphasize patterns.}
    \label{fig:ex:airports}
\end{figure*}

Following Ferreira et al., we compare trips departing from Lower Manhattan to LaGuardia and JFK airports over the first week of May 2011. \autoref{fig:ex:airports} shows the corresponding directional queries (\textit{a} and \textit{b}) and the patterns revealed in the embedded time series plots. Note how the number of trips varies over the week (with a large spike on Friday afternoon, \autoref{fig:ex:airports}--\textit{c}) and how the trip duration in minutes changes according to the day of the week and time of the day (\autoref{fig:ex:airports}--\textit{d}). Meanwhile, the trip distances naturally remain constant, as do the fares per trip (\autoref{fig:ex:airports}--\textit{e}), since trips from Manhattan to airports have predetermined prices. As in this scenario fewer trips are selected, connecting lines (\autoref{fig:ex:airports}--\textit{f}) can be used to emphasize their patterns. 
The opposite query is also possible, selecting trips starting at a transportation hub and observing the most common destinations in the city, as shown in \autoref{fig:egocentric}--\textit{left} with Grand Central Station. For this, both the on-demand 2D map projections and 2D choropleth mode are useful. 

Switching to the egocentric room-scale perspective allows the analyst to step onto the map and walk around to observe points of interest more closely (\autoref{fig:egocentric}--\textit{left}). In some cases, the egocentric 360º images from \textit{Google StreetView} can also aid in the exploration of potential causes for large numbers of taxi destinations in a specific area (or the lack thereof). \autoref{fig:egocentric}--\textit{right}, for example, reveals a hotel in Manhattan---not surprisingly, the picture itself captures a large number of yellow taxis. In a collaborative exploration scenario, it is possible to imagine two analysts doing such inspections together or exploring different perspectives simultaneously.

\subsection{Using comparative brushing and embedded attribute plots to investigate activity in different city regions}

One of the ideas behind \textit{TaxiVis} is using taxis as urban sensors which can help to reveal broader urban behaviors. One such possible analysis is to compare taxi activity in different regions of the city. Although the same analysis is also possible with visual queries, \autoref{fig:ex:neighborhoods} shows a comparative exploratory analysis between different neighborhoods using both pointer brushes (whose selections immediately affect all coordinated plots) and the choropleth mode brushing that builds prism stacks evidencing periods of higher activity. 

Another example, \autoref{fig:teaser}--\textit{center right}, demonstrates how the use of recurrent queries helps the user to focus on specific times. In this case, we investigate incoming taxi activity (dropoffs) in Lower Manhattan in the evenings (6PM-12AM). The sliced prisms emphasize the currently selected times, while the point cloud and embedded plots show that activity grows over the week (upwards, in this case), and the on-demand choropleth map specifies the number of trips that originated at other Manhattan neighborhoods. 

Building on these examples, we can conceive multiple application scenarios that incorporate other \textit{immersive extensions}. In a collaborative scenario, multiple users could quickly investigate different hypotheses through appropriate extensions. For example, the query duplication button helps to create multiple variations of a query targeting different areas or times. The future integration of private and shared views could thus be useful to enable parallel exploration of different aspects. In a casual exploration scenario, a general citizen could be interested in exploring the activities in their neighborhoods, or compare their personal experiences to the city trends. In a real-time data scenario, query prisms could emphasize the currently applied filters when only some trips are relevant.

\section{Discussion}

Having designed, implemented, and demonstrated our proof-of-concept, we revisit our original question on the potential benefits of IA for urban data exploration in this section. We discuss our current perspective on benefits and drawbacks of the immersive paradigm, technical challenges, and our research vision for the next years.

\subsection{Researchers' perspective: What does the immersive paradigm bring to the table?}

Our team includes researchers from both Urban Visualization and IA backgrounds, and with this project, our goal was to pursue a critical reflection, highlighting potential benefits but also noting potential drawbacks of an immersive approach to urban analytics. 

The following initial collected perceptions correspond to our opinions based on our experience developing the \textit{Immersive TaxiVis} system, and thus should not be interpreted as a completely unbiased validation. Still, we believe some potential benefits are clear. As illustrated in \autoref{sec:demo:hurricanes}, some analyses can benefit from a better visualization of the temporal component of the data, which is well supported in a 3D STC view. Without a VR interface, such an analysis would lack sufficient depth cues and rely on (effortful) 3D manipulation through 2D interaction devices, or would involve extremely cluttered 2D visualizations. Using the third dimension also allowed us to better integrate and correlate trip attribute data, juxtaposing embedded time series with queries and brushes, as shown in \autoref{fig:ex:sandy} and \autoref{fig:ex:airports}, for example. Prior IA work, such as \textit{ImAxes} \cite{cordeil2017imaxes, cunningham2021geospatial}, also enabled intuitive exploration of combinations of attributes, and this seems to be a promising direction. In the future, allowing direct transformations between 2D and 3D views should also further increase their usefulness \cite{lee2022transformations,seraji2022hybridaxes}.

In a world of growing familiarity with spatial interfaces, the use of intuitive metaphors such as sculpting queries \cite{hurter2018fiberclay} and simultaneously brushing origin and destination by two-handed interaction can potentially simplify data exploration activities, conceivably decreasing the system-incurred cognitive load and making tools more accessible to people of different backgrounds. As also demonstrated by \textit{Uplift} \cite{ens2021uplift}, such tools can facilitate knowledge sharing between different kinds of stakeholders. In a single-user scenario, the use of a VR HMD is also potentially helpful to enable more focused exploration, removing external distractions. We also foresee the growing use of VR applications for user engagement, particularly in the context of motivating casual data exploration by citizens.

We also consider that our 3D prisms clearly achieve the purpose of indicating the applied space-time filters accurately and enabling direct modifications to them. We see the decomposition of complex interfaces into spatial components connected to 3D objects as another appropriate simplification that leverages humans' abilities to interact with spatial environments. Being able to efficiently interact with the 3D terrain and building models is also an advantage within many urban analysis scenarios where the data is connected to the local topography or urban density. Finally, recent work also clearly indicates the potential for immersive collaboration \cite{lee2021shared,saffo2023shoes}.

\subsection{Current limitations of immersive urban analytics}

On the other hand, we also identified limitations and concerns, especially for the short-term future. While we believe \textit{Immersive TaxiVis} supports exploratory data analysis well, the most efficient placement of some interface components is still not entirely clear. For example, when experimenting with complex queries combining many atomic ones, we observed that the process could become more complex than in a conventional interface, where keyboard shortcuts facilitate multiple selections, a functionality that \textit{Immersive TaxiVis} currently does not support. The risk of potentially over-complicating the exploration through the limitations of the user interface is present and must be navigated by the designers, though we expect that best practices will emerge. Conversely, we also see the risk of over-simplifying, seeking to merely reproduce desktop workflows (or even only parts of them). The idea of immersive interfaces should be to extend and complement desktop ones and, in this context, it is not surprising that not all tasks will be simpler under this paradigm. 

Other current concerns include the disproportionate difficulty of developing IA systems, particularly ones that can be deployed to real applications. While the often-used Unity platform provides a vast set of resources, including dealing with maps, data rendering, and interaction toolkits, their integration is not straightforward, and some, such as \textit{IATK}, still pose a considerable learning curve for more advanced applications. This challenge is even greater in the case of systems that must scale to very large datasets. Another concern is the limited generalizability of existing prototypes to other datasets, as many approaches, including in our case, are partly customized.

\subsection{Implementation challenges and scalability}
\label{sec:disc:impl}

The main challenges we faced in building our prototype correspond to identifying the most adequate resources for implementing each component, leading to constant rework; learning to extend the functionality of \textit{IATK}; and finding solutions that would scale to large urban datasets without compromising the high frame rate required.

As originally discussed in the \textit{IATK} paper \cite{cordeil2019iatk}, the toolkit is able to sustain good performance up to around two million data points, after which performance drops significantly. 
Our implementation for spatio-temporal querying and brushing, however, sustains a very satisfactory interactive performance for datasets up to a hundred thousand trips (that is, two hundred thousand points, or six hundred thousand points with all projections enabled). Above that size, query modifications can cause the system to freeze for a few seconds, partially hindering the immersive experience. This is mostly due to the cost of computing query stats for the associated visualizations. 
As reference: we worked in PC-VR mode, with the Oculus Quest 2 headset directly connected to a desktop PC equipped with an i7-8700 3.2GHz CPU, nVidia GTX 1070 GPU, and 16GB of RAM.

While our current proof-of-concept system already efficiently supports many moderately-sized urban datasets, for example, those obtained from origin-destination surveys \cite{menin2019estime}, others, such as the taxi and bike-sharing datasets from NYC can only be currently imported in subsets, as they contain millions of points per week or month. We intend to continue optimizing query computations to support larger datasets and will assess options to decouple the query computing and rendering components in the future.

\subsection{Research vision and the path ahead}

\subsubsection{For immersive urban analytics}

We expect that, through better support for context-switching between immersive and non-immersive approaches, most of the issues discussed above will be either addressed or significantly reduced over the next years, enabling users to benefit from the best of both ``worlds''. More capable devices will also allow users to more seamlessly switch between systems to access different interaction modalities, e.g., text entry on a keyboard and 3D manipulation through a controller. More accurate and precise interaction devices, together with enhanced display resolutions, will also support efficient 2D interaction within immersive systems. Moreover, asymmetric collaboration, where one user investigates the data within an immersed perspective while another interacts with a more conventional interface, can also overcome some of the limitations \cite{saffo2023shoes, tong2023asymmetric}. 

From an engineering perspective, we expect that novel toolkits will simplify the development of IA applications, for example, through the use of visualization grammars \cite{lee2023deimos}. Web-based urban visualization toolkits \cite{moreira2023urban,uber2018kepler} will likely also be increasingly adapted into AR and VR through \textit{WebXR} platforms. 

In terms of applications, we expect to see the development of novel approaches to take advantage of more photo-realistic 3D city models, for example, to directly visualize the effects of buildings in the urban space \cite{ferreira2015urbane,zhang2021urbanvr}. As immersive applications are already seeing increasing adoption in the context of architecture \cite{delgado2020architecture}, an extension to urban planning is thus also expected. In this context, the use of immersive systems will also likely be applied to inform and engage citizens in the planning process \cite{schrom2020planning, keil2023measuring}. As a medium-term goal, the integration of multiple sources of urban data, including real-time data, will certainly be of interest to many applications.

\subsubsection{For the \textit{Immersive TaxiVis} system}

In future work, our goal is to complement \textit{Immersive TaxiVis} with additional functionality, such as 2D and 3D heat map visualizations, and \textit{immersive extensions}. In particular, we will look at collaborative analysis activities performed by multiple users. We will also further refine the design of our user interface and consider options to improve the system's performance to support larger urban datasets.

While we see \textit{Immersive TaxiVis} as a proof-of-concept demonstration for the \textit{immersive extensions} and \textit{adaptations} we envisioned, we also consider it a great platform for evaluations with users. In future work, we thus intend to conduct expert and usability evaluations using different urban datasets. While this was beyond the scope of this paper, such evaluations will be essential to better understand how data exploration workflows vary between different paradigms and how they can complement each other. Towards this purpose, we plan to improve the generalizability of our visualizations to support additional datasets, for example, extending views in the attribute exploration panels to also support categorical attributes.

\section{Conclusion}

Above all, this work was a reflection on how the growing adoption of immersive paradigms for data visualization can contribute to the visualization of urban datasets from the point of view of Immersive Analytics and Urban Visualization research. Our reflections were guided by the development of an immersive proof-of-concept prototype for the seminal \textit{TaxiVis} system in urban data visualization. We designed our prototype with the simultaneous intentions of preserving key functionality of the original desktop interface while also complementing that set of features with new ones enabled by \textit{immersive extensions}. With a series of examples related to the NYC taxi dataset, we demonstrated how this system could be used in practice, for example, to more clearly inspect behaviors over time or to inspect locations of interest with complementary egocentric perspectives. We envision possible continuations of this work through expert and usability user evaluations and by reimagining additional well-known 2D data exploration systems as a way of identifying additional potential benefits of IA environments.

\acknowledgments{\textit{Jorge Wagner} was supported in this project by the Fulbright Foreign Student Program, by a Microsoft Research PhD Fellowship, and by the Brazilian National Council for Scientific and Technological Development (CNPq). \textit{Claudio Silva} is partially supported by the DARPA PTG program, National Science Foundation CNS-1828576, and NASA. Any opinions, findings, conclusions, or recommendations expressed in this material are those of the authors and do not necessarily reflect the views of DARPA. 
\textit{Wolfgang Stuerzlinger} acknowledges support from NSERC.
\textit{Luciana Nedel} also acknowledges support from CNPq (311259/2022-7,  421962/2023-2). This study was financed in part by the Coordenação de Aperfeiçoamento de Pessoal de Nível Superior - Brasil (CAPES) - Finance Code 001.}

\appendix
\section{Supplementary Materials: Notes on the prototype system implementation}
\label{sec:taxivis:implementation}

We implemented our proof-of-concept prototype using the Unity game engine, the most frequently used platform in IA research, given the multitude of valuable toolkits available for it. In this section, we highlight relevant resources we used and key implementation decisions we took, as well as their present limitations, with the goal of informing subsequent projects.

\subsection{Data rendering resources}

To overcome Unity's performance limitations that arise with the creation of a large number of scene objects, we decided early on to adopt the \textit{Immersive Analytics Toolkit} (\textit{IATK}) \cite{cordeil2019iatk}. To our knowledge, \textit{IATK} is presently the most efficient form of implementing big data visualization in Unity, being a more accessible alternative than developing a custom rendering engine, as done, for example, by \textit{FiberClay} \cite{hurter2018fiberclay}. 

\textit{IATK}'s efficiency is the result of a series of clever design decisions: in \textit{IATK}, data points are not represented as individual scene objects, but as vertices in a single big mesh. Depending on the selected visualization topology, different Unity shaders (i.e., code that runs on the computer's GPU) render this mesh in the appropriate way. 
Correspondingly, interactions with data points are not detected by individual colliders, but by the intersection of brushes with mesh vertices, efficiently computed by Unity \texttt{Compute Shader} programs and encoded into the RGB pixel values of two-dimensional textures with size $\sqrt{n} \times \sqrt{n}$, where $n$ is the number of points in the dataset.

However, the main limitation of \textit{IATK} is that it was mostly intended for manual configuration of views through the Unity editor interface, with limited and little-documented support for code-based customization. 
We believe our approach to adapting it to suit our specific needs can serve as an example for future IA applications.

\subsection{Interaction and mapping resources}

In terms of user interaction, we decided to adopt the \textit{Mixed Reality Toolkit} (\textit{MRTK}). This toolkit provides excellent support for easy implementation of interface components, with support for multiple devices of different brands through the \textit{OpenXR} protocol and good support for integration with the \textit{Oculus Package} to support both controller and hand-tracking actions as enabled by the \textit{Oculus Quest 2} HMD, the one we worked with in this project. Future works may also consider other alternatives, such as the \textit{Unity XRI Toolkit}.

To develop our STC visualization, we adopted the \textit{Bing Maps SDK} for Unity to integrate a dynamic mapping API. This SDK provides efficient rendering of 3D terrain meshes and also of 3D buildings for some cities, including New York City, as seen in \autoref{fig:teaser}--\textit{center}. Several other alternatives are also available and may be preferable depending on the project, including \textit{Cesium for Unity} using photo-realistic 3D data from \textit{Google Maps}, the \textit{Mapbox Maps SDK for Unity}, or the \textit{ArcGIS Maps SDK for Unity}.

\subsection{Integrating STC queries into IATK}

Our prototype implementation, whose source code is publicly available, demonstrates how the \textit{IATK} code and its core optimizations can be programmatically adapted to construct a query-able STC. We adapted the original \texttt{Scatterplot View} from \textit{IATK} to be dynamically attached to the interactions performed on the STC map and time walls. For datasets with a bipartite nature such as taxi data (independent OD pairs), we create two \texttt{View} components with \texttt{Point} topologies, as well as a \texttt{LinkingView}, whose visual OD connections are hidden by default but can be displayed on demand. We also create four additional \texttt{View} components to render bidimensional projections on space and time when enabled.

To implement efficient display and querying of millions of data points, \textit{TaxiVis'} architecture decoupled its rendering and storage components. Replicating this behavior would require dynamically generating new \textit{IATK} big meshes for each query result, which could be impracticable without hindering interactivity and breaking the immersive experience. Therefore, for our proof-of-concept, we opted to instead heavily adapt the original \texttt{BrushingAndLinking} \textit{IATK} class to compute all queries through Unity shaders and encode query results into three texture images (encoding filtered points, brushed points, and highlighted points). We also significantly extended the \textit{IATK} \texttt{MyComputeShader} to enable filtering by arbitrary 3D prisms, and to enable several such prisms to be combined, through texture post-processing done by a second \texttt{ComputeShader} program, depending on the intended query logic. The overall query logic is managed by our \texttt{QueryManager} class.

By coupling the rendering and querying components of the system, we gain the ability to efficiently apply spatio-temporal filters that scale up to hundreds of thousands of data points, without having to regenerate the \textit{IATK} \texttt{View} mesh. At the same time, this prevents us from computing queries on millions of data points, something that \textit{TaxiVis} achieved (technically speaking, \textit{TaxiVis} never rendered more than one million points at the same time, but could query and compute stats over larger sets of points in the background). We see this as a reasonable implementation compromise at this time.

Given our performance concerns to maintain a comfortable frame rate for VR exploration, we also worked to reduce the number of view updates when possible. 
Further technical details about our query implementation approach are beyond the scope of this publication, but are documented in our GitHub repository.

\bibliographystyle{IEEE/abbrv-doi-hyperref-narrow}

\bibliography{main}

\end{document}